# Proposal for combined use of parallax and lunar laser ranging for Michelson-Morley experimentation

**Akinbo Ojo**
Standard Science Centre
*P.O. Box 3501, Surulere, Lagos, Nigeria*
taojo@hotmail.com

**Abstract**
The null findings of the terrestrially conducted Michelson-Morley type experiments have been influential to determining which physical theory best fits reality. Here, we propose the use of parallax and ranging techniques of distance measurement for Michelson-Morley type experimentation on astronomical scales, elaborating with the earth-moon system. A motivation is the bypass of any hypothetical earth-bound medium as an explanation for the earlier null findings. When the solar system 370 kms$^{-1}$ motion relative to the cosmic microwave background is taken into consideration, the higher value of receptor velocity and the longer distance sets the stage for a confrontation between Lorentzian and Galileo-Newton dynamical transformations. Two-way ranging time carries along with it information about light's dynamical transformation, thus a comparison of distance measured by ranging with that by parallax, a geometrical method can reveal the correctness of assumptions underlying the dynamical behaviour of light. Two-way ranging times unaffected by the 370 kms$^{-1}$ velocity indicate null results and further buttress Einstein's relativity theory, while ranging times affected by this velocity indicate non-null results and will manifest as a general increase of ranging distance above actual geometric lunar distance, up to 500 metres in RA 11h 11.649m and 23h 11.649m coordinate directions.

**Key words:** parallax, lunar laser ranging, special relativity, SLR

PACS Classification: 03.30.+p; 11.30.Cp; 95.10.Jk

## I. Introduction

Michelson-Morley type experiments [1] have been fundamental to the development of physical theory and they always will be. The motivation was to find out if the earthly motion of a receptor towards or away from an incoming light beam would have any effect on the arrival time of light. The effect, if present was expected to show by an interference pattern, hence the naming of the light path as an interferometer. Several versions of the experiment have been conducted all indicating that the earthly motion of the receptor has no significant effect on the arrival time of a photon in transit. This null finding, "… the unsuccessful attempts to discover the effect of the motion of the earth relatively to the light medium" [2] was elevated to a universal principle (Lorentz invariance) in the formulation of Einstein's theory of relativity [2,3].

To explain the obedience to the discovered Lorentzian principle, physical mechanisms were needed and were proposed. For a receptor moving away from an incoming light beam, the path length is proposed to be contracted by the required amount necessary to prevent a delayed arrival time so that the resultant light velocity is retained at *c*. This is the FitzGerald- Lorentz contraction mechanism. For a receptor moving towards an incoming light beam, a time dilation effect explained the preservation of an arrival time compatible with a resultant velocity, *c*.

In the Michelson-Morley experiment, both the source and receptor were earth-bound and stationary to the earth, leading to alternative suggestions that perhaps a hypothetical earth-bound medium which may be ether [4] or a dark matter [5] was responsible for the null experimental results.

Continuing controversy spurred other experiments, notably that of Sagnac [6]. Using rotational motion, he was able to demonstrate that the arrival time of light to a receptor can be influenced by the motion of the receptor relative to the earth. This further fueled the hypothesis that the earlier null results could be due to an earth-bound medium, which when the receptor was able to move relative to the medium, a relative light velocity could be detected. However being non-linear, the Sagnac results have not been unanimously accepted as a violation of Lorentz invariance and explanations have been made for it within General relativity. Recent modified Sagnac experiments [7] using uniform motion relative to the earth seem to find non-null results, again resuscitating the possibility of an earth-bound medium.

The astronomical value of light velocity makes dynamical experiments difficult to test the proposals concerning the constancy of the one way speed of light because firstly, this requires the receptor to have a high enough velocity relative to that of light and secondly, a sufficiently long path length is needed to make light arrival time differences significant. Both are difficult to achieve terrestrially for experimenting the Special relativity (SR) and Galileo-Newton (GN) velocity addition laws. Other difficulties arise in geometrically measuring the suggested displacement contraction required to ensure the constancy of the one way speed of light because the measuring instrument is proposed to also contract by the same small amount as the path length.

Timing technology at better than micro-second accuracy and increasing precision of long distance measurement by non-dynamical methods have become available. The design of experiments whereby a receptor can move at high velocity towards and away from an incoming light beam travelling over an astronomical path length therefore appears to be a useful proposition. One advantage of such an experiment is that it bypasses any hypothetical earth-bound medium as explanation for any null results that may be obtained.

In this paper, we discuss our proposal how a combined use of parallax and ranging techniques may be useful for the conduct of Michelson-Morley type experiments on an astronomical scale. Significantly, parallax and ranging techniques determine distance by independent methods. While the parallax method uses geometrical considerations, ranging is based on the dynamical behaviour of light. A combination of both, subject to the required accuracy levels therefore presents an opportunity to test and discuss the correctness of the various proposals on light's dynamical transformation.

We organize the paper by firstly, describing the parallax and ranging methods in section II. We then take note of the basic facts about the earth-moon system in section III and in section IV describe the features and arrangement that could make the system amenable to conduct a Michelson-Morley type investigation. In section V, we look at possible results that might be obtained and discuss them in section VI. We end with concluding remarks in section VII.

## II. Earth-moon distance measurement

(a) Parallax method.

The moon being the most visible of celestial objects from earth was the earliest to be subjected to distance measurement. Hipparchus, 190 - 120 B.C., is recorded as making the earliest successful attempt. The technique used was the parallax method which involves measuring the angle subtended due to apparent change in position of a near object when viewed from different locations against a distant background. Knowing the angle and the distance between observing positions, the distance to the object can be calculated, see Fig.1. This method is based on geometry. The precision of the measurement depends on how accurately the distance between the observing positions, A and B is known and the precision of measured angles.

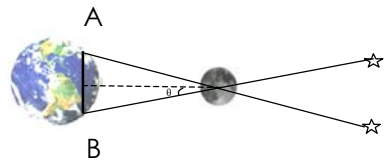

Fig.1. Showing the basis for parallax measurement of distance.

The basic geometrical equation is Eq.(1).

$$\text{distance}_{moon} = \text{half of distance}_{observer\ base} / \text{tangent of angle } \theta \qquad (1)$$

The use of photography to take pictures of the moon at exactly the same time using synchronized clocks from two locations on earth and comparing the apparent change in position relative to the distant stars is now routinely employed by observing teams in determining the distance to the moon, e.g. see [8].

(b) Ranging method.

This involves observing the two-way transit time, $\tau$ of a laser or radar signal transmitted to an object and reflected back. Knowing this transit time and light velocity, $c$, the distance to the object can be calculated. This is a dynamical method. Following the successful placing of reflectors on the lunar surface, the measurement of distance between an observatory on earth and a laser reflector on the lunar surface has provided several lunar laser ranging (LLR) measurements. Many of these are publicly available online as Normal Point data [9].

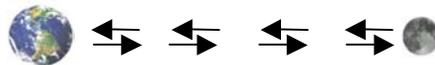

Fig.2. Illustrating the Lunar Laser Ranging method.

The basic observation equation for LLR is Eq.(2) and has been used to determine earth-moon distance to high accuracy [10,11].

$$d = c\tau/2 = ( r_{em} - r_{station} + r_{reflector} ) + c\Delta\tau \qquad (2)$$

where d is the station-reflector distance, $c$ is the speed of light, $\tau$ is the two-way travel pulse time, $r_{em}$ is the vector connecting the geocenter and selenocenter, $r_{station}$ is the geocentric position vector of the observatory and $r_{reflector}$ is the selenocentric position vector of the reflector. $\Delta\tau$ describes corrections of the travel time caused by atmospheric effects, Shapiro delay and other relativistic corrections.

To calculate the distance, Eq.(2) does not accord much importance to the linear velocity of the receptor relative to the incoming light while it is on its way. This is in keeping with the Lorentzian principle that the velocity of light and thus arrival time of an incoming light beam cannot be affected by the receptor's own velocity. However Eq.(2) takes cognizance of the effect the rotational velocities of the observing station and reflector can have due to a rotating earth and moon. It also takes into account the changing earth-moon distance due to the elliptical orbit.

Having tacitly incorporated Lorentz invariance into Eq.(2), it may appear that LLR cannot be useful for testing alternative transformations. However, obtained timing results are the observational data and only the distance deduced depends on assumptions about the dynamical transformation applicable to light. Obtained ranging time and the distance calculated from it, when placed alongside distance measurements by a non-dynamical method can therefore reveal the correctness of the underlying assumptions by finding either an agreement or discrepancies between dynamically calculated measurements of distance using Eq.(2) and the geometrically obtained distance from parallax. It is important to state here that if discrepancies are found, this must be statistically significant, within error limits and be better explained by alternative dynamical assumptions.

### III. Basic facts of the earth-moon system

Astronomical observations and data provide the following relevant features:

- The moon orbits the earth at a mean distance ~ 384,403 km or about 1.28 light-seconds. The orbit is nearest to earth at perigee (~ 363,104 km) and furthest at apogee (~ 405,696 km) and inclined 18.3° to 28.6° to the earth's equatorial plane.

- The earth revolves around the sun at an average speed of 30 kms$^{-1}$.

- The earth rotates around the sun counter-clockwise as seen from the north (prograde motion) and the moon orbits the earth also counter-clockwise as seen from the north.

- The sun itself is in motion, (along with the earth-moon system). Relative to the cosmic microwave background, the sun is moving 370 kms$^{-1}$ towards a point with approximate galactic coordinates, $l = 264°.4$, $b = 48°.4$ (RA = 11h 11.649m, $\delta$ = -7° 11.947'), near the Virgo constellation [12].

To put this diagrammatically, see Fig.3.

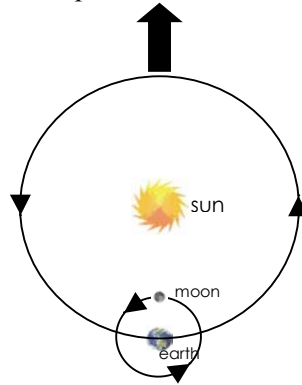

370 kms$^{-1}$ motion in space towards RA 11h 11.649m, δ -7° 11.947'

Fig.3. Showing the motion of the sun, earth and moon in space.

## IV. Experimental set up

The necessary experimental arrangement for a Michelson-Morley type investigation are (i) a light source, S (ii) a light receptor, R and (iii) a light path of known distance capable of being oriented in such a way that the receptor (observer) can at various times be moving towards or away from the incoming light. This is illustrated in Fig.4. The idea is to find out whether the motion of the receptor relative to the incoming light has any effect on light's arrival time.

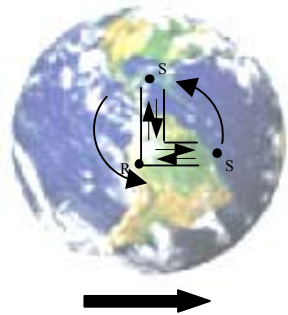

Motion of earth-based source and receptor

Fig.4. A simplified representation of the Michelson-Morley apparatus.

Although one-way experimentation between a source and a receptor would be the ideal way to investigate Lorentz invariance, results can be muddled in clock synchronization disputes especially over long distances, i.e. "when exactly was the received light emitted at a distant source?" or "when exactly did the outgoing signal arrive a distant receptor?" Less glaring differences may be present, but two-way experiments avoid such clock synchronization problems as only one clock is required.

We next briefly describe our proposal how the earth-moon system can be used for a Michelson-Morley type investigation. Apart from timing technology, the necessary parts of the apparatus are:

(i) Light source: An earth-based observatory where laser beams can be sent to the moon.

(ii) Light receptors: One on the moon capable of reflecting and thus verifying laser beam arrival and another in the earth-based observatory capable of receiving reflected laser light.

(iii) Light path: The light path in the original Michelson-Morley experiment is the interferometer arm moving in space along with the earth. The interferometer arms are measured geometrically to ensure the equality of light paths to be compared. Analogous to the interferometer arm in the terrestrial version, we propose a similarly geometrically measured light path, which will be the earth observatory and moon reflector distance as measured by parallax. See similarity between Figs. 4 and 5.

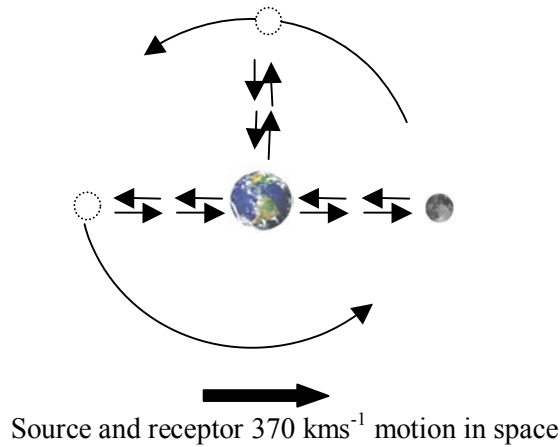

Source and receptor 370 kms$^{-1}$ motion in space

Fig.5. A representation of the basics for an earth-moon Michelson-Morley investigation.

A combination of parallax determinations of distance before an outgoing laser pulse is sent, during transit and on arrival back at the observatory and the two-way transit time recordings made at different orbital positions taking into consideration the 370 kms$^{-1}$ earth-moon motion in space will provide the effect of source and receptor moving away and towards incoming light and this can be investigated. Fig.6 shows the positions of most observational interest.

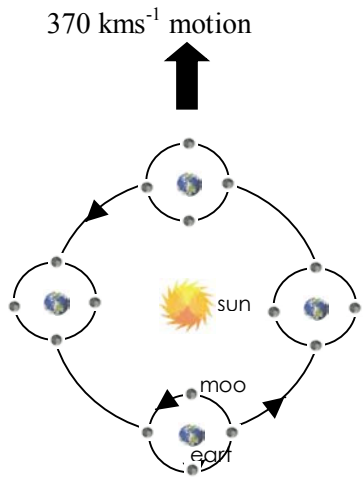

Fig.6. Showing observational positions of most interest.

## V. Possible results

A. Constancy of the one-way speed of light.

A first possible result is that laser beam arrival times are unaffected by the motion of the earth or moon away from or towards incoming light. The resultant velocity of light also always remains conserved at $c$ by means of time dilation and displacement contraction, such that the effective path length travelled remains essentially unchanged between the source and receptor during light transit. This is the Lorentz transformation.

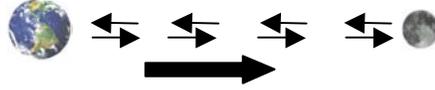

Fig.7. Illustrating LLR and the 370 kms$^{-1}$ earth-moon motion present in certain positions.

For example, giving a parallax determined distance, d of 384,000 km and light travelling at velocity, $c$ equal to $3 \times 10^8$ ms$^{-1}$, two-way transit time, $\tau$ will be 2.56 seconds because even while the laser beam is in transit, motion of the moon away from or towards it cannot delay or hasten its arrival time and similarly when it is incoming, motion of the earth towards or away can also not hasten or delay its arrival time. Ranging distance, $c\tau/2$, always equals the geometrical distance, d.

B. Variability of the one-way speed of light.

In SR, the velocity of the receptor, v always amounts to zero so that addition of velocities always gives $c$ and d always equals $c\tau/2$. However in the GN view, velocity of light can be relative to that of the receptor and so v can be positive when the receptor is travelling towards the incoming light and negative if travelling away from it. This makes light arrive earlier or later depending on the receptor's velocity towards it while the laser beam is in transit.

For a two-way ranging where the receptor's velocity is directly away or towards the incoming laser beam, the two-way travel time, $\tau$ in the GN transformation will be

$$\tau = d/(c-v) + d/(c+v) \qquad (3)$$

where v is the velocity of the receptor towards the incoming light travelling at velocity, $c$ and d is the geometric distance. Taking the expression for $\tau$ in Eq.(3) into account, the two-way ranging equation can therefore be written

$$d = (c\tau/2)(1 - v^2/c^2) = (r_{em} - r_{station} + r_{reflector})(1 - v^2/c^2) + c\Delta\tau \qquad (4)$$

where the term, $1 - v^2/c^2$ is incorporated to allow for the velocity of the receptor relative to the incoming light. For a Lorentz transformation, v = 0 and this recovers Eq.(2). Eq.(4) therefore appears as the ranging equation that can accommodate both the Lorentz and GN transformation possibilities.

Observationally, positions where the 370 kms$^{-1}$ velocity of the receptor relative to the incoming light is maximal, i.e. 6 and 12 o'clock, should reveal the most effect on arrival times, if this is present. In some orbital positions, the experimental arrangement may also be useful to investigate the relative importance of the earth's orbital velocity, e (30 kms$^{-1}$) to the solar system motion, s (370 kms$^{-1}$). Practically, where s applies e becomes relatively insignificant being comparatively smaller.

Taking the velocity addition possibility of GN into account, the following results can be looked for during observations at the positions shown in Fig.6.

(i) Earth at 6 o'clock position

Moon at 6 o'clock  $\tau = d/(c+s)$ [outgoing] + $d/(c-s)$ [incoming]

  3 o'clock  $\tau = d/(c-e)$ [outgoing] + $d/(c+e)$ [incoming]

  12 o'clock  $\tau = d/(c-s)$ [outgoing] + $d/(c+s)$ [incoming]

  9 o'clock  $\tau = d/(c+e)$ [outgoing] + $d/(c-e)$ [incoming]

(ii) Earth at 3 o'clock position

Moon at 6 o'clock  $\tau = d/(c+s+e)$ [outgoing] + $d/(c-s-e)$ [incoming]

  3 o'clock  $\tau = d/c$ [outgoing] + $d/c$ [incoming]

  12 o'clock  $\tau = d/(c-s-e)$ [outgoing] + $d/(c+s+e)$ [incoming]

  9 o'clock  $\tau = d/c$ [outgoing] + $d/c$ [incoming]

(iii) Earth at 12 o'clock position

Moon at 6 o'clock  $\tau = d/(c+s)$ [outgoing] + $d/(c-s)$ [incoming]

  3 o'clock  $\tau = d/(c+e)$ [outgoing] + $d/(c-e)$ [incoming]

  12 o'clock  $\tau = d/(c-s)$ [outgoing] + $d/(c+s)$ [incoming]

  9 o'clock  $\tau = d/(c-e)$ [outgoing] + $d/(c+e)$ [incoming]

(iv) Earth at 9 o'clock position

Moon at 6 o'clock  $\tau = d/(c+s-e)$ [outgoing] + $d/(c-s+e)$ [incoming]

  3 o'clock  $\tau = d/c$ [outgoing] + $d/c$ [incoming]

  12 o'clock  $\tau = d/(c-s+e)$ [outgoing] + $d/(c+s-e)$ [incoming]

  9 o'clock  $\tau = d/c$ [outgoing] + $d/c$ [incoming]

The receptor's velocity, v can be s or e or a combination of both as spelt out above.

To illustrate the GN velocity addition law, for light travelling a parallax determined distance, d, e.g. 384,000 km to a receptor moving towards it with velocity, s, e.g. with earth at 6 o'clock, moon at 6 o'clock, the one-way travel time, t will be

t = d/ (c+s) = 1.278423278 seconds

On the return journey, the incoming light will be travelling in the opposite direction with the receptor moving away, so travel time will be

t = d/ (c-s) = 1.281580616 seconds

This brings the total two-way travel time, $\tau$ to 2.560003894 seconds. For this two-way time, a ranging distance based on Lorentzian assumptions will give a geometric distance 584 m longer than the actual 380,000 km distance.

For a GN transformation, if observed two-way ranging time, $\tau$ is 2.56 seconds, this will correspond to a geometric distance of 383,999.416 km, which is shorter than what the distance would be were the transformation Lorentzian. Thus for the same two-way ranging time, the Lorentzian transformation yields a longer geometric distance than the GN transformation.

We see from Eq.(4), that if contrary to SR assumptions, the velocity of the receptor is indeed relevant and the transformation in LLR is GN, a fraction of the ranging distance equal to $v^2/c^2$ has to be deducted from the ranging distance, $c\tau/2$ to obtain the correct geometric distance, d.

Geometric distances as measured by parallax will therefore always equal the ranging distance within experimental error, if observed two-way time is SR rather than GN. However if the observed two-way ranging time is GN but wrongly assumed to be Lorentzian, the actual geometric distance will always be shorter than the ranging distance by an amount equal to a fraction ($v^2/c^2$) of the ranging distance. The more suitable dynamical transformation can therefore be looked for by comparing the ranging and parallax distance measurements.

For orthogonal positions, i.e. where the laser beam is transmitted in a position orthogonal to the moon's 370 kms$^{-1}$ velocity, e.g. at 3 and 9 o'clock positions, the transformation is slightly different, see Fig.8.

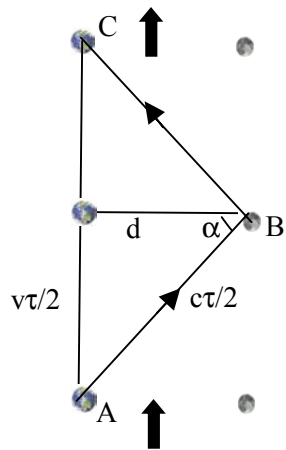

Fig.8. Showing earth-moon 370 kms$^{-1}$ motion in bold arrows and the laser beam path.

When the 370 kms$^{-1}$ motion of the earth and moon is taken into consideration, a laser beam sent from the earth at position A with the moon adjacent, reaches the moon when it must have moved to B and returns back to the observatory which must now be at C.

If the transformation is Lorentzian and the 370 kms$^{-1}$ receptor velocity does not count, the geometric distance, d will as usual equal the ranging distance, $c\tau/2$.

If however, the transformation is GN and the earth-moon system 370 kms$^{-1}$ velocity counts as shown, then from Pythagoras theorem,

$$d^2 = (c\tau/2)^2 - (v\tau/2)^2 \qquad (5)$$

Putting $c/c$ in the term $v\tau/2$, we obtain

$$d = (c\tau/2)(1 - v^2/c^2)^{1/2} \qquad (6)$$

For an observed two-way ranging time of 2.56 seconds, the moon would still be geometrically 384,000 km away in a Lorentzian transformation but 383,999.708 km away in a GN transformation for positions orthogonal to the 370 kms$^{-1}$ motion. This is a difference of about 292 m which can be checked by parallax. In summary, for positions orthogonal to the earth-moon 370 kms$^{-1}$ velocity, if the ranging transformation is GN rather than Lorentzian as assumed, a fraction ($v^2/c^2$) of the ranging distance must be deducted from its square to obtain the square of correct geometric distance.

## VI. Discussion

The earth-moon distance is analogous to the interferometer arm in the terrestrially conducted Michelson-Morley type experiment. Both can be rotated and oriented such that incoming light can travel in different directions relative to the earth's motion in space.

Atmospheric and gravitational effects will apply. These are not mentioned but being usually small compared to transit times and being common to both SR and GN the effects may not make differentiation difficult. In the Lorentzian view, effects of gravity are described with relative time, while in the GN perspective, it is light velocity that may be affected by gravity and not time. Both can be expressed with slight modification by the Shapiro delay. The advantage of this proposed experimental arrangement includes eliminating any possible earth-bound medium as an explanation for the null results of the terrestrially conducted versions, if the findings here are also null.

If the transformation applicable in LLR is SR, distances calculated from obtained timing results using Eq.(2) will always equal that observed geometrically by parallax within error limits. However, if the transformation applicable in ranging is GN rather than SR, unless the appropriate terms are deducted from the ranging distance, distances from obtained timing data will always exceed the actual geometric distance and the difference can be as much as 584 metres.

The proposed experiment is in the main limited by accuracy of the parallax method. The 1" seeing limits constrains the possibility to go below an error of 1/3600, which is about 0.03% (~ 100 km). This is not good enough for detection of micro-light second differences in distance. The use of adaptive optics, more sophisticated telescopes and making ranging measurements and parallax from the same location with multiple baselines may possibly help in reducing parallax error, including errors involved with reduction of results to earth centre. It is noteworthy however that the lunar ephemerides published by the US Naval Observatory in the *Astronomical Almanac* ( http://aa.usno.navy.mil) claims a precision for horizontal parallax up to 0.0003" (~ 32 m). Ranging data reduced to centre-to-centre distance between the earth and moon can be compared to this.

Using Astronomy software such as Starry Night Pro the approximate dates for the positions of most interest can be found for past and future years. For the year 2010 A.D., the following dates apply:

(i) Earth will be at 12 o'clock position (RA = 11h 11.649m ) ~ 7th March
Moon will be at 6 o'clock     ~ March 15
              3 o'clock     ~ February 23
             12 o'clock   ~ March 1
              9 o'clock     ~ March 8

(ii) Earth will be at 9 o'clock position (RA = 17h 11.649m ) ~ 9th June
Moon will be at 6 o'clock     ~ June 5
              3 o'clock     ~ June 12
             12 o'clock   ~ June 18
              9 o'clock     ~ May 29

(iii) Earth will be at 6 o'clock position (RA = 23h 11.649m) ~ 9th September
Moon will be at 6 o'clock     ~ September 22
              3 o'clock     ~ September 2
             12 o'clock   ~ September 8
              9 o'clock     ~ September 15

(iv) For Earth at 3 o'clock position (RA = 5h 11.649m) ~ 11th December
Moon will be at 6 o'clock     ~ December 13
              3 o'clock     ~ December 20
             12 o'clock   ~ November 29
              9 o'clock     ~ December 6

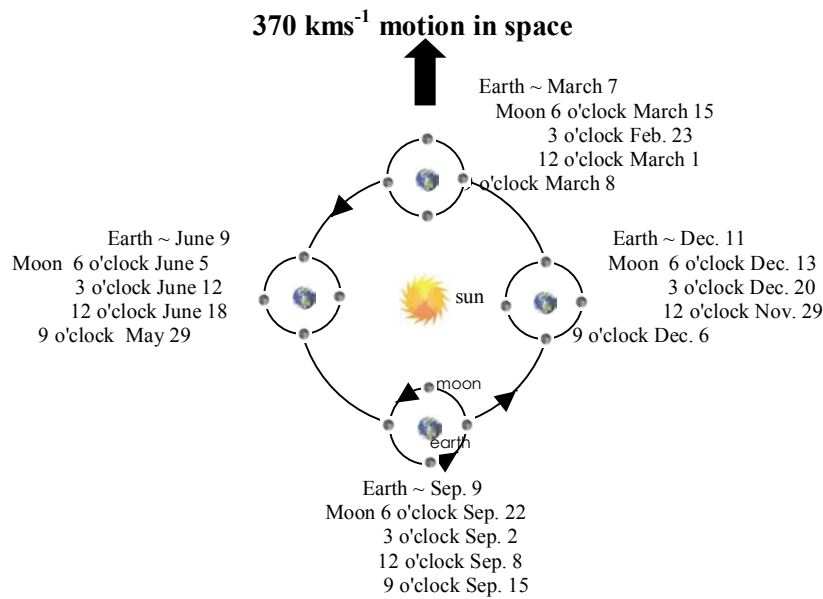

Fig.9. Showing approximate dates for observational positions of most interest (2010).

At the moment, the Apache Point Observatory Lunar Laser-ranging Operation, (APOLLO) claims the most ranging accuracy and their data are publicly available at http://www.physics.ucsd.edu/~tmurphy/apollo/norm_pts.html. For the different earth positional dates in Fig.9, parallax and ranging measurements around each date will capture the lunar cycle and positions corresponding to 3, 6, 9 and 12 o'clock. The percentage differences between LLR and parallax can thereby be available for comparison. Null results would indicate the universality of Lorentz invariance, while non-null findings suggest the opposite.

Other phenomena that may be looked out for include aberration at the 3 and 9 o'clock positions. There appears to be a possibility of more signal loss in those positions because the 370 kms$^{-1}$ motion of the moon and earth is orthogonal to the laser beam direction. If present, reducing photon loss might involve pointing the laser beam appropriately by an angle α towards position B in Fig.8 when the earth is at A and again for best reception seconds later when the earth would be at C. A rough estimate of angle α shows it will be very small, about 4.24', probably making it of theoretical interest only.

Parallax measurements and radar ranging to Venus and Mars provide other possible arrangements for similar experimentation, particularly when the planets are at superior conjunction and the light path aligned with the solar system 370 kms$^{-1}$ motion. Being more distant than the moon, differences between SR and GN can be more obvious.

In summary, with improved accuracy of parallax measurement, the following can be buttressed or refuted experimentally:

- The solar system or orbital velocity of an earth-bound observer has no effect on the resultant velocity of light. Therefore the null result of the terrestrial Michelson-Morley experiment is not due to an earth-bound medium but a consequence of a Lorentzian principle which seems to be more universal.

- If a resultant velocity of light due to solar system motion is observed, this can be quantified and can serve as an independent method of determining the magnitude and direction of the earth's absolute velocity in space for comparison with the dipole anisotropy observations of Doppler shifts in CMB [12].

- If a resultant velocity due to the earth's motion is observed, the non-null result will need to be placed alongside the null result of the terrestrial Michelson-Morley experiment for proper interpretation.

**VII. Concluding remarks**

The 370 kms$^{-1}$ solar system motion in space presents interesting scenarios for both the Galileo-Newton (GN) and Special relativity (SR) dynamical transformations. For the GN transformation, this higher value of the earthly motion through space, being more than the 30 kms$^{-1}$ that was modestly being looked for in the terrestrial Michelson-Morley experiment, implies that the results of that experiment were at least ten times even further indicative of a nullity. For SR, the higher values of receptor velocity and the longer lunar distance provide a laboratory for further testing the findings of the terrestrial Michelson-Morley experiment and the adequacy of the mechanisms proposed to explain the null results such as displacement contraction and time dilation.

Ranging data have been mostly used to research General relativity and not Special relativity, even though two-way ranging time carries along with it information about light's dynamical transformation. Unlike parallax, the technique has reached such sophistication as to tempt the assertion that the accuracy level of distance measurement based on obtained timing data now approaches ~ 2 cm. Such confidence must however be corroborated by a non-dynamical method such as parallax. As has been demonstrated here, if the transformation in ranging is Newtonian, rather than Lorentzian as is currently assumed, lunar distances claimed by LLR will be generally higher and can be as far off the mark from actual geometric distances with as much as ~ 470 km (one-way), ~584 m (two-way in the coordinate directions RA 11h 11.649m and 23h 11.649m) or ~292 m for orthogonal positions. Until the parallax method has therefore been used to establish the more appropriate dynamical transformation, it is prudent for a *caveat* to be placed on distances measured by ranging in the development of accurate lunar ephemerides.

Collaboration between teams separately working on parallax and ranging can refine what has been proposed here. If the technical challenges can be overcome, the results of a Michelson-Morley type investigation on an astronomical scale as described here should go some way in unifying our views about the dynamical behaviour and nature of light.

**Acknowledgements**
I thank Davide Cenadelli for comments about the intricacies of the parallax method with which he is more familiar.